\documentclass[floatfix,aps,amsmath,amssymb,reprint]{revtex4-2}

\usepackage{graphicx}
\usepackage{dcolumn}
\usepackage{bm}
\usepackage{placeins}
\usepackage{float}
\begin{document}

\title{Morphology of active deformable 3D droplets}%

\author{Liam J. Ruske}
\email{liam.ruske@physics.ox.ac.uk}
\author{Julia M. Yeomans}
\affiliation{Rudolf Peierls Centre For Theoretical Physics, University of Oxford, UK}

\date{\today}

\begin{abstract}
We numerically investigate the morphology and disclination line dynamics of active nematic droplets in three dimensions. Although our model only incorporates the simplest possible form of achiral active stress, active nematic droplets display an unprecedented range of complex morphologies. For extensile activity finger-like protrusions grow at points where disclination lines intersect the droplet surface. For contractile activity, however, the activity field drives cup-shaped droplet invagination, run-and-tumble motion or the formation of surface wrinkles. This diversity of behaviour is explained in terms of an interplay between active anchoring, active flows and the dynamics of the motile dislocation lines. We discuss our findings in the light of biological processes such as morphogenesis, collective cancer invasion and the shape control of biomembranes, suggesting that some biological systems may share the same underlying mechanisms as active nematic droplets.\\ \\
\end{abstract}

\maketitle

\section{\label{sec:intro}Introduction}

Active particles use energy from their surroundings to do work. Examples range from eukaryotic cells, bacterial suspensions and motor proteins to active colloids and shaken granular rods \citep{narayan2007long, duclos2014perfect}. Active systems have recently received considerable attention in the community because of their potential in designing mesoscopic engines, as a way of interpreting biological mechanics, and as examples of non-equilibrium statistical physics \citep{doostmohammadi2018active, marchetti2013hydrodynamics}. Here we focus on dense active systems, and in particular the continuum theory of active nematics, which describes active systems with hydrodynamic interactions. The archetypal experimental example is microtubules driven by kinesin motor proteins \citep{zhang2016dynamic, keber2014topology}. Other active nematics include myosin-driven actin-microtubule networks \citep{lee2020myosin}, swimming bacterial swarms and confluent eukaryotic cells \citep{saw2017topological, mueller2019emergence, duclos2017topological, saw2018biological}.

A key property of active nematics, which distinguishes them from passive liquid crystals, is active turbulence. This is a chaotic flow state characterised by strong vorticity  and motile topological defects which are continually created and destroyed. Considerable experimental and theoretical work has been devoted to understanding the properties of active turbulence, and the associated topological defects, in two dimensions. More recently it has proved possible to design an active material that allows imaging of a three-dimensional active nematic and, in particular the associated motile disclination loops and lines \citep{duclos2020topological}. This was achieved by dispersing force-generating microtubule bundles in a passive colloidal liquid crystal based on filamentous viruses. The temporal evolution of disclination lines was measured using light-sheet microscopy revealing that the primary topological excitations in bulk active nematics are charge-neutral disclination loops. 

Numerical simulations have revealed that flows and morphological dynamics of disclination lines in three dimensional (3D) active nematics are governed by the local director profile surrounding the disclination line and that defect loops in extensile systems are generally formed via the well-known bend instability \citep{binysh2020three}. Three-dimensional active nematic turbulence and disclination line dynamics in spherical confinement has also been investigated using numerical modeling in both achiral \citep{vcopar2019topology} and chiral systems \citep{carenza2019rotation, carenza2020chaotic}. Enforcing strong in-plane surface alignment allowed the formation of defects on the surface and was used to highlight the coupling of surface and bulk topological defect dynamics.

The theories of active materials are increasingly being used to describe biological systems in two dimensions, with examples including biofilm initiation \citep{yaman2019emergence}, topological defects in cell monolayers \citep{duclos2017topological,saw2017topological,mueller2019emergence} and epithelial expansion \citep{blanch2017hydrodynamic}. This suggests that in three dimensions there may be relevance to the collective motion of groups of cells, in morphogenesis or to the growth and spread of tumours. Therefore to underpin extension of these approaches to 3D, in this paper we investigate active, self-deforming droplets in three dimensions, describing the interplay between disclination dynamics and droplet morphology. We find that extensile droplets form protrusions at points where disclination lines reach the surface. In contractile droplets, however, dislocations lead to a wrinkled drop surface. Moreover for small, contractile droplets, we find that invagination, to form a cup-like configuration, can be driven by nematic activity alone. We present evidence explaining the reasons behind the different behaviours stressing, in particular, the role of active anchoring, an effective surface alignment resulting from active flows.

We start by presenting the mathematical description of an active nematic and the equations of motion which we solve numerically, and follow this by a summary of 3D active nematics and the classification of disclination lines. The main results  are presented in section IV, which is split into four subsections: First we investigate the nematic structure on the surface of an active droplet (A) and how this affects the dynamics of disclination lines in the bulk of spherical droplets (B). We then move on to show how active forces can deform the interface by highlighting mechanisms causing droplet deformations triggered by both extensile (C) and contractile (D) activity. The last section of the paper summarises the key results and points out possible connections to biological systems.

\section{\label{sec:theory}Equations of motion}

We summarise the continuum equations of motion describing the dynamics of a deformable, nematic droplet in an isotropic fluid background. In the absence of activity the system is described by a free energy $\mathcal{F} = \int f \: d\mathbf{V} = \int f_{GL} + f_{LC} \: d\mathbf{V}$. The first contribution, which controls the formation of the nematic droplet, is chosen to take the Ginzburg-Landau form
\begin{equation}
	f_{GL} = \frac{\kappa^*}{2} \left( -\varphi + \varphi^3 - \epsilon^2 \mathbf{\nabla}^2 \varphi \right)^2 + \frac{K_{\varphi}}{2} \left( \mathbf{\nabla} \varphi \right)^{2} \: .
\end{equation}
This describes phase separation into two stable phases with concentrations $\varphi=\pm1$ with an interface of width $\epsilon$ separating the two phases~\citep{lazaro2015phase, helfrich1973elastic}. The bending rigidity $\kappa$ of the interface is related to $\kappa^*$ by $\kappa^*=(4 \epsilon^3 / 3 \sqrt{2}) \kappa$. The final term penalizes gradients in the concentration field and introduces a surface tension $\sigma \propto \sqrt{K_{\varphi}}$. 

The second contribution to the free energy density is 
\begin{multline}
	f_{LC} = A_{LC} \Bigg\{ \frac{1}{2} \left( 1-\frac{\eta(\varphi)}{3}\right) \:  \mbox{tr}(\mathbf{Q}^{2}) - \frac{\eta(\varphi)}{3} \:  \mbox{tr}(\mathbf{Q}^{3})  \\
	+ \frac{\eta(\varphi)}{4} \: \mbox{tr}(\mathbf{Q}^{2})^{2}  \Bigg\} + \frac{1}{2} K_{LC} \left( \mathbf{\nabla} \mathbf{Q} \right)^{2} \: 
\end{multline}
which includes the usual Landau-de Gennes bulk energy of the liquid crystal expanded in terms of the nematic tensor $\mathbf{Q}$ which has elements $Q_{ij}=3 S_0 /2(n_i n_j - \delta_{ij}/3)$, where $S_0$ is the magnitude and $\mathbf{n}$ the direction of the nematic order, and a term which penalizes elastic deformations of the director field \citep{de1993physics}. $A_{LC}$ sets the bulk energy scale and we use the one-elastic-constant approximation which assigns the same elastic constant $K_{LC}$ to twist, splay and  bend deformations. The function $\eta(\varphi)$, which quantifies the dependence of bulk energy on the the local liquid crystal concentration predicts a first order phase transition between a nematic and an isotropic phase of the liquid crystal at the value $\eta_{c}=2.7$. We define $\eta(\phi) := \eta_{c} + \eta_{1} \varphi$, and choose a value of  $\eta_{1}$ that ensures that the system forms a nematic droplet ($\varphi=+1$) which is surrounded by isotropic fluid ($\varphi=-1$).

Since the total concentration $\int \varphi dV$ is assumed to stay constant, diffusive transport follows Model-B dynamics and the time evolution of $\varphi$ is governed by the following reaction-diffusion equation \citep{cahn1958free}:
\begin{equation}
	\left( \partial_{t} + \mathbf{u} \cdot \mathbf{\nabla} \right) \varphi = \Gamma_{\varphi} \: \mathbf{\nabla}^{2} \mu\:.
	\label{EoM_phi}
\end{equation}
Here $\mathbf{u}$ is the velocity field and the mobility $\Gamma_{\varphi}$ quantifies how fast $\varphi$ responds to gradients in the chemical potential $\mu=\delta \mathcal{F}/\delta \varphi$.

Unlike the total concentration, the local nematic alignment $\mathbf{Q}$ is not a conserved quantity and its time evolution follows modified Model A dynamics \citep{beris1994thermodynamics}:
\begin{equation}
	\left( \partial_{t} + \mathbf{u} \cdot \mathbf{\nabla} \right) \mathbf{Q} - \mathbf{S} = \Gamma_{Q} \mathbf{H} , \:
	\label{EoM_Q}
\end{equation}
where $\Gamma_{Q}$ is the rotational diffusivity and  $\mathbf{H}$, the molecular field, is given by
\begin{equation}
\mathbf{H}=-\left[ \frac{\delta \mathcal{F}}{\delta \mathbf{Q}} - \frac{1}{3} \: \mathbf{I} \: tr \left( \frac{\delta \mathcal{F}}{\delta \mathbf{Q}} \right) \right] \: .
\end{equation}
Rod-like particles can not only be advected by the fluid, but also rotate in response to flow gradients. This behaviour is accounted for by the co-rotational term \citep{beris1994thermodynamics} %
\begin{multline}
	S_{ij} = \left( \xi D_{ik}+\Omega_{ik} \right) \left( Q_{kj} + \frac{\delta_{kj}}{3} \right) + \left( Q_{ik} + \frac{\delta_{ik}}{3} \right) \\
	\left( \xi D_{kj}-\Omega_{kj} \right) - 2\xi \left( Q_{ij} + \frac{\delta_{ij}}{3} \right) Q_{kl} W_{lk}, \:
\end{multline}
where $D_{ij}=(\partial_j u_i + \partial_i u_j)/2$ and $\Omega_{ij}=(\partial_j u_i - \partial_i u_j)/2$ are the symmetric and antisymmetric parts of the velocity gradient tensor $W_{ij}=\partial_i u_j$, respectively. The parameter $\xi$, determines whether the director  aligns with, or tumbles in, a shear flow.
Its value depends on the molecular details of the liquid crystal, and it is related to the flow alignment parameter $\lambda$ defined within Ericksen-Leslie theory by $\lambda = \frac{3 S_0 +4}{9 S_0} \xi$. This work mostly focuses on flow tumbling nematics, for which the flow alignment parameter $\lambda<1$.

We assume that the build-up of active flows occurs over time scales which are much longer than any viscoelastic relaxation time of the system. We therefore consider the liquid limit and solve the incompressible Navier-Stokes equations to obtain the flow field $\mathbf{u}$:
 \begin{equation}
 \nabla \cdot \mathbf{u} = 0 \: ,
 \label{EoM_compress}
 \end{equation}
\begin{equation}
	\rho \left( \partial_{t} + \mathbf{u} \cdot \mathbf{\nabla} \right) \mathbf{u} = \mathbf{\nabla} \cdot \Pi , \:
	\label{EoM_fluid}
\end{equation}
where the stress tensor $\Pi = \Pi_{viscous} + \Pi_{elastic} + \Pi_{capillary} + \Pi_{active}$. The passive contributions, well known from liquid crystal hydrodynamics \citep{marenduzzo2007hydrodynamics}, are:
\begin{equation}
 \Pi_{viscous} = 2 \eta \mathbf{D} \:,
\end{equation}
 \begin{multline}
\Pi_{capillary} = (f-\mu \varphi) \mathbf{I} - \mathbf{\nabla} \varphi \left( \frac{\partial f}{\partial (\mathbf{\nabla} \varphi)} \right) \\
 +\mathbf{\nabla} \varphi \nabla \left( \frac{\partial f}{\partial (\mathbf{\nabla}^2 \varphi)} \right) - \mathbf{\nabla} \mathbf{\nabla} \varphi \left( \frac{\partial f}{\partial (\mathbf{\nabla}^2 \varphi)} \right)\: ,
 \end{multline}
\begin{multline}
 \Pi_{elastic} = -p \mathbf{I} - \xi [ \mathbf{H} \Tilde{\mathbf{Q}} + \Tilde{\mathbf{Q}} \mathbf{H} - 2 \Tilde{\mathbf{Q}}  tr(\mathbf{Q} \mathbf{H}) ] + \mathbf{Q} \mathbf{H} \\
 - \mathbf{H} \mathbf{Q} - \mathbf{\nabla} \mathbf{Q} \left( \frac{\partial f}{\partial (\mathbf{\nabla} \mathbf{Q})} \right) \: , \\
 \end{multline}
where $\rho$ is the density, $\eta$ the viscosity, $p$ the bulk pressure and $\Tilde{\mathbf{Q}} = \left(\mathbf{Q}+\frac{1}{3} \mathbf{I}\right)$. In addition to the passive terms, the stress due to the dipolar flow fields produced by the active particles is \citep{simha2002hydrodynamic}
\begin{equation}
	\Pi_{active} = - \zeta \mathbf{Q} \: ,
\end{equation}
where $\zeta$ quantifies the magnitude of active stress. For extensile activity, $\zeta>0$, nematic particles direct the fluid outwards along their elongated direction and inwards along the two perpendicular axes. The flow direction is reversed for contractile activity, $\zeta<0$.

We solved the equations of motion using a hybrid lattice Boltzmann-finite difference method. This involved solving eqs.~(\ref{EoM_phi}, \ref{EoM_Q}) using finite difference methods, and eqs.~(\ref{EoM_compress}, \ref{EoM_fluid}) using a lattice Boltzmann algorithm \citep{marenduzzo2007steady,de1993physics,beris1994thermodynamics,doostmohammadi2016defect}. Simulations were performed on a three-dimensional lattice of size 100x100x100 and discrete space and time steps were chosen as unity. We used periodic boundary conditions for the simulation box and the system initially consisted of a spherical, nematic drop of radius $R=30$ ($\varphi=+1$) embedded in an isotropic fluid ($\varphi=-1$), unless otherwise stated. Initially we let the system relax for $t=500$ time-steps such that the droplet interface and the nematic tensor $\mathbf{Q}$ reached their equilibrium profile before we switched on activity. We used the following parameter set for all simulations: $\rho=1$, $p=0.25$, $\Gamma_Q=0.1$,  $\Gamma_\varphi=0.2$, $\kappa^*=0.1$, $\epsilon=1.4$, $A_\varphi=0.1$, $\eta_1=0.3$, $\xi=0.1$ and $\eta=1/3$ in lattice-Boltzmann units. For spherical droplets in sections IV.A and IV.B, we used $K_{LC}=0.1$, $K_\varphi = 0.4$, $A_{LC}=0.75$ and $|\zeta|=0.01$. For soft droplets in sections IV.C and IV.D, we used the following parameters, unless otherwise stated: $K_{LC}=0.2$, $K_\varphi = 0.2$, $A_{LC}=1.5$,and $|\zeta|=0.02$.\\

\section{Disclination lines in three-dimensional active nematics}

\begin{figure}
	\centering
	\includegraphics[width=8.6cm]{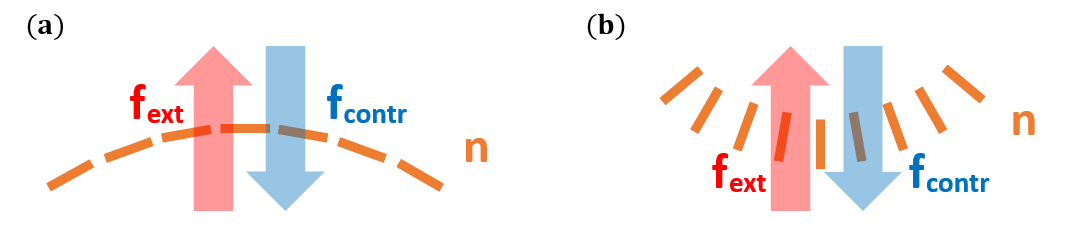}
	\caption{Bend (a) and splay (b) perturbations of the director field $\mathbf{n}$ cause active forces which set up active flows which further enhance or stabilise the respective perturbation. Forces in extensile or contractile systems are denoted by red or blue arrows, respectively. It is apparent that extensile (contractile) systems are unstable to bend (splay) deformations.}
	\label{fig_hydr_inst}
\end{figure}

Active stress leads to an instability of the nematic phase (Fig.~\ref{fig_hydr_inst}). This hydrodynamic instability constantly pushes the system out of equilibrium, leading to a chaotic steady state termed active turbulence \citep{thampi2016active}. 3D active turbulence is characterised by spatiotemporally chaotic flows and the presence of disclination lines which constantly undergo transformation events such as breakup, recombination, nucleation and annihilation \citep{ vcopar2019topology}. In bulk systems disclination lines typically form closed, charge-neutral loops \citep{duclos2020topological}. However, in the presence of a confining interface, as for nematic droplets, disclination lines can also terminate at the boundary and the dynamics of the resultant defects on the surface is coupled to disclination line dynamics in the bulk by elastic interactions and flows.

\begin{figure}
	\centering
	\includegraphics[width=8.6cm]{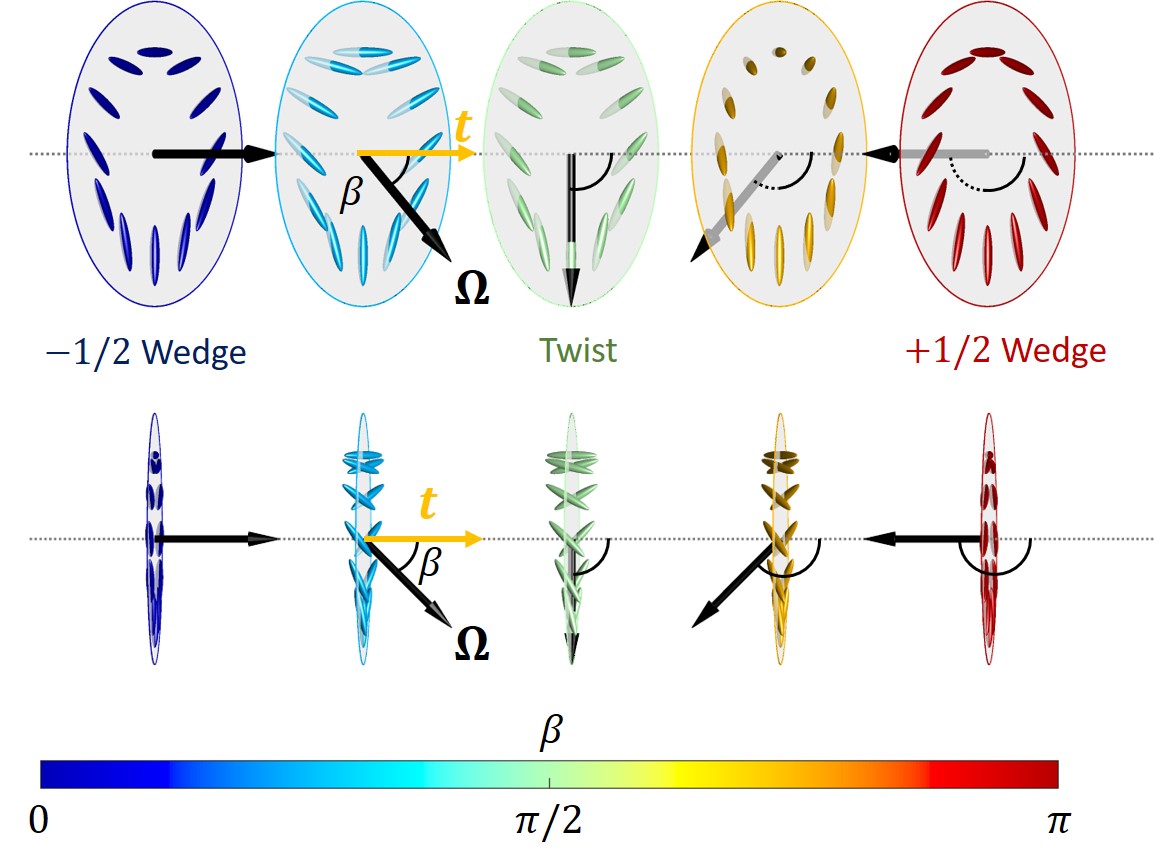}
	\caption{A local $-1/2$ wedge can continuously transforms into a $+1/2$ wedge via an intermediate twist disclination \citep{duclos2020topological}. The director field winds about the rotation vector $\mathbf{\Omega}$ (black arrows) by an angle $\pi$. The angle between the rotation vector $\mathbf{\Omega}$ and the local disclination line tangent $\mathbf{t}$ (yellow arrow), called the {\it twist angle} $\beta$, varies continuously along a disclination line. Twist disclinations correspond to $\beta = \pi/2$ and $+1/2$ and $-1/2$ wedge disclinations to $\beta = \pi$ and $0$, respectively.}
	\label{fig_beta_sketch}
\end{figure}

Unlike in two-dimensional active nematics where $\pm 1/2$ defects carry topological charge and can therefore only nucleate or annihilate in pairs of opposite charge, disclination lines in three dimensions can continuously transform from a local $-1/2$ configuration (in the plane perpendicular to the line) into a $+1/2$ configuration through an intermediate twist winding as indicated in Fig.~\ref{fig_beta_sketch}. As one moves around the core of a disclination in the plane perpendicular to the local disclination line segment, the director field winds around a specific axis, the rotation vector $\mathbf{\Omega}$ (black arrows) by an angle $\pi$. The angle $\beta$ between $\mathbf{\Omega}$ and the local line tangent $\mathbf{t}$ (yellow arrow) is called the {\it twist angle} and can be used to locally characterise the disclination line. 
For $-1/2$ ($+1/2$) wedge-type defects the twist angle corresponds to $\beta=0 (\pi)$, while line segments with local twist-type defects are indicated by $\beta = \pi/2$.

Due to activity disclination lines act as self-propelled entities moving through the fluid. Based on a simplified model neglecting elastic interactions, each disclination line segment can be associated with a local self-propulsion velocity which depends on the local director profile \citep{binysh2020three}. The self-propulsion velocity component perpendicular to the local tangent of the line depends on twist angle $\beta$ as
\begin{equation}
    v^{SP}_{\perp} \propto (1-\cos{\beta})^2 \: .
    \label{eqn_spv}
\end{equation}
Thus, unlike in two-dimensional active turbulence where the dynamics is dominated by two species of quasi-particles ($\pm 1/2$-defects), disclination lines  in three-dimensional active turbulence act as quasi-particles with an intrinsic degree of freedom, the twist angle $0\leq \beta \leq \pi$. 
Line segments with $\beta=0$  are passive while $\beta=\pi$ line segments are most active in pushing around the surrounding fluid.

\section{Results}

\subsection{\label{sec:actalgn}Activity leads to preferred director alignment at an interface}

In 2D active nematics, the flow induced by active stresses leads to an alignment of the director at an interface with an isotropic phase \citep{blow2014biphasic}. This {\it active anchoring} of the director field is parallel to the interface for extensile stress and perpendicular for contractile stress. For 2D systems, the active alignment can be significant, but it is not immediately obvious whether the effect will persist in a 3D geometry, or how it will be affected by any defects present on the surface.\\

\begin{figure}
	\centering
    \includegraphics[width=8.6cm]{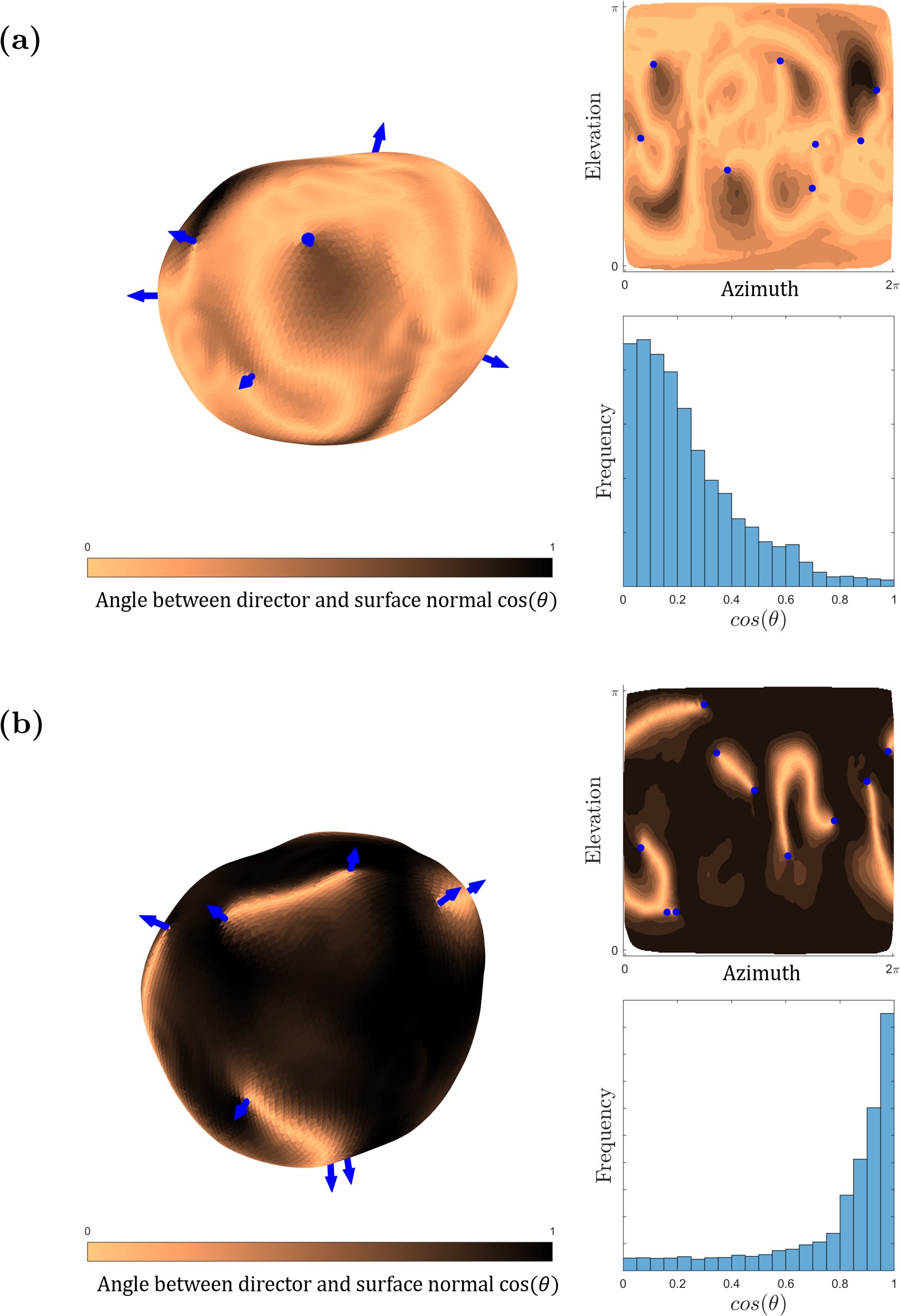}
	\caption{Surface alignment of the director field for extensile (a) and contractile activity (b). Surface defects, shown as blue arrows, are connected via disclination lines running through the bulk. The surface alignment is indicated by colour bar, where orange (black) indicates in-plane (perpendicular) director alignment with respect to the interface. Upper right: surface alignment in projection. Lower right: distribution of surface angle $\cos(\theta)$ over the total surface area}
	\label{fig_act_anch}
\end{figure}

Therefore, to investigate the effects of active anchoring in 3D, we measure the angle $\theta$ between the director field $\mathbf{n}$ and the surface normal at the interface of spherical droplets. Since there are more possible configurations for in-plane alignment ($\theta=\pi/2$) than for perpendicular alignment ($\theta=0$), we use the distribution of $\cos{\theta}$ to quantify alignment effects. A randomly aligned director field results in a uniformly distributed $\cos{\theta} \sim \mathcal{U}[0,1]$, while preferred perpendicular or in-plane alignment leads to bias of the distribution towards $1$ or $0$, respectively. The results in Fig.~\ref{fig_act_anch} show clear evidence that  both extensile and contractile activity lead to strong active anchoring on the surface, which weakly depends on the flow-alignment parameter $\lambda$ (Fig.~S1 \citep{si}).

\begin{figure}
	\centering
	\includegraphics[width=8.6cm]{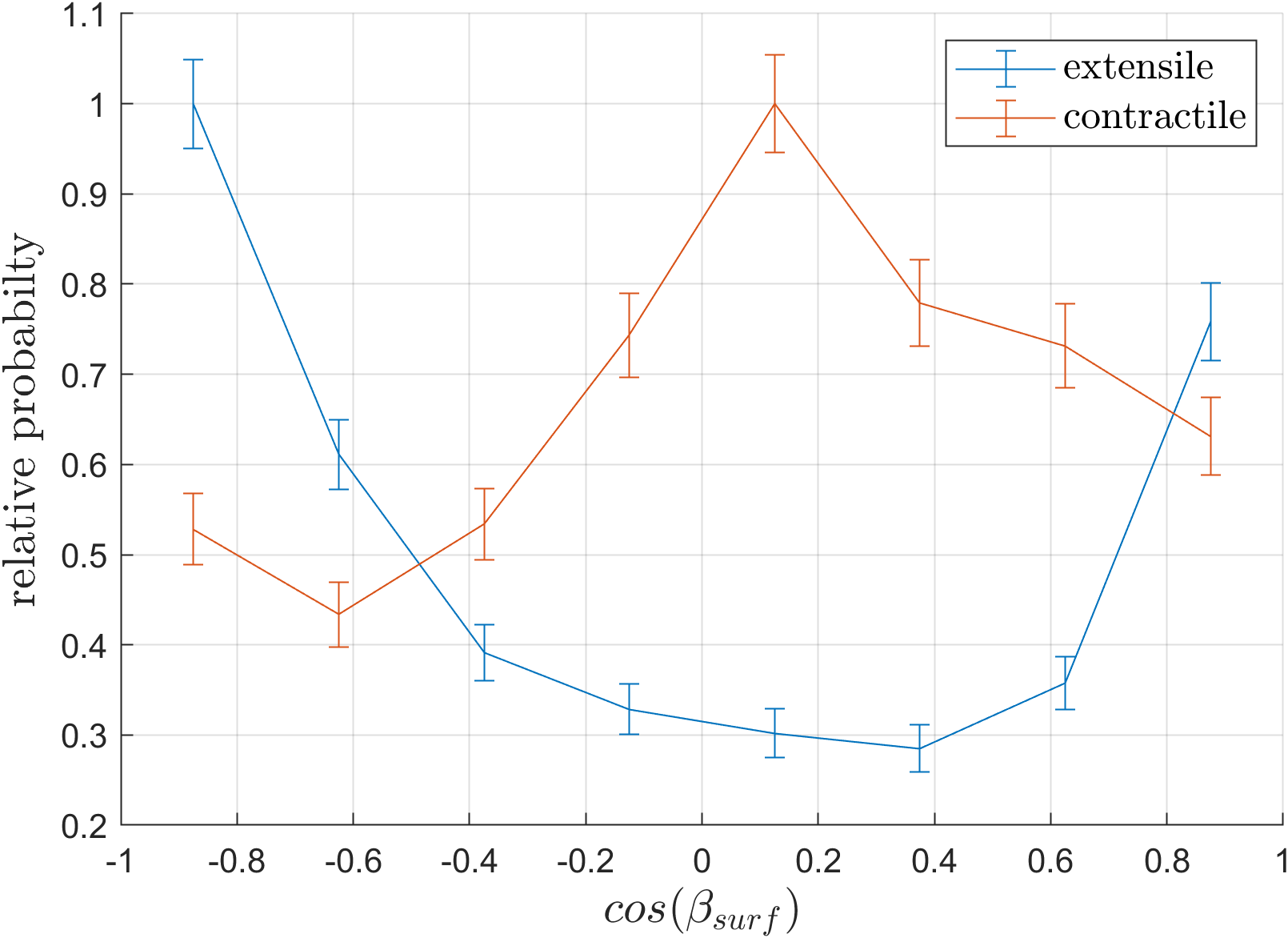}
	\caption{Distribution of twist angles of surface defects  $\beta_{surf}$, i.e. at positions where disclination lines intersect the droplet interface, obtained from a time average in the active turbulent regime. Surface defects are preferentially wedge-like (twist-like) in extensile (contractile) systems where active anchoring favours in-plane (normal) director alignment.}
	\label{fig_betasurf}
\end{figure}

The director orientation also shows signatures of the places where  disclination lines end at the surface. Consider first the case of extensile activity. The director lies in plane over most of the surface, although there  are localised regions with perpendicular alignment (see Fig.~\ref{fig_act_anch}(a), Movie~S1 \citep{si}). The disclination lines present in the bulk of the droplet create quasi two-dimensional defects at the positions where they terminate at the interface which we shall term surface defects. The distribution of the twist angle $\beta$ of surface defects has distinct peaks at $\beta=0$ and $\beta = \pi$, showing that most defects on the surface are of wedge type (Fig.~\ref{fig_betasurf}), corresponding approximately to 2D +1/2 and -1/2 configurations. There are less twist-type surface defects with $\beta \approx \pi/2$  as these introduce a perpendicularly aligned region in the vicinity of the defect, which is suppressed by active anchoring.

A two-dimensional nematic sheet confined to the surface of a sphere always has at least four $+1/2$ defects present as the total topological charge is conserved and must add up to two, the Euler-characteristic $\chi$ of a sphere. Active anchoring does not confine the director field to the surface everywhere so the topological charge is not strictly conserved. \\

\begin{figure}
	\centering
	\includegraphics[width=8.6cm]{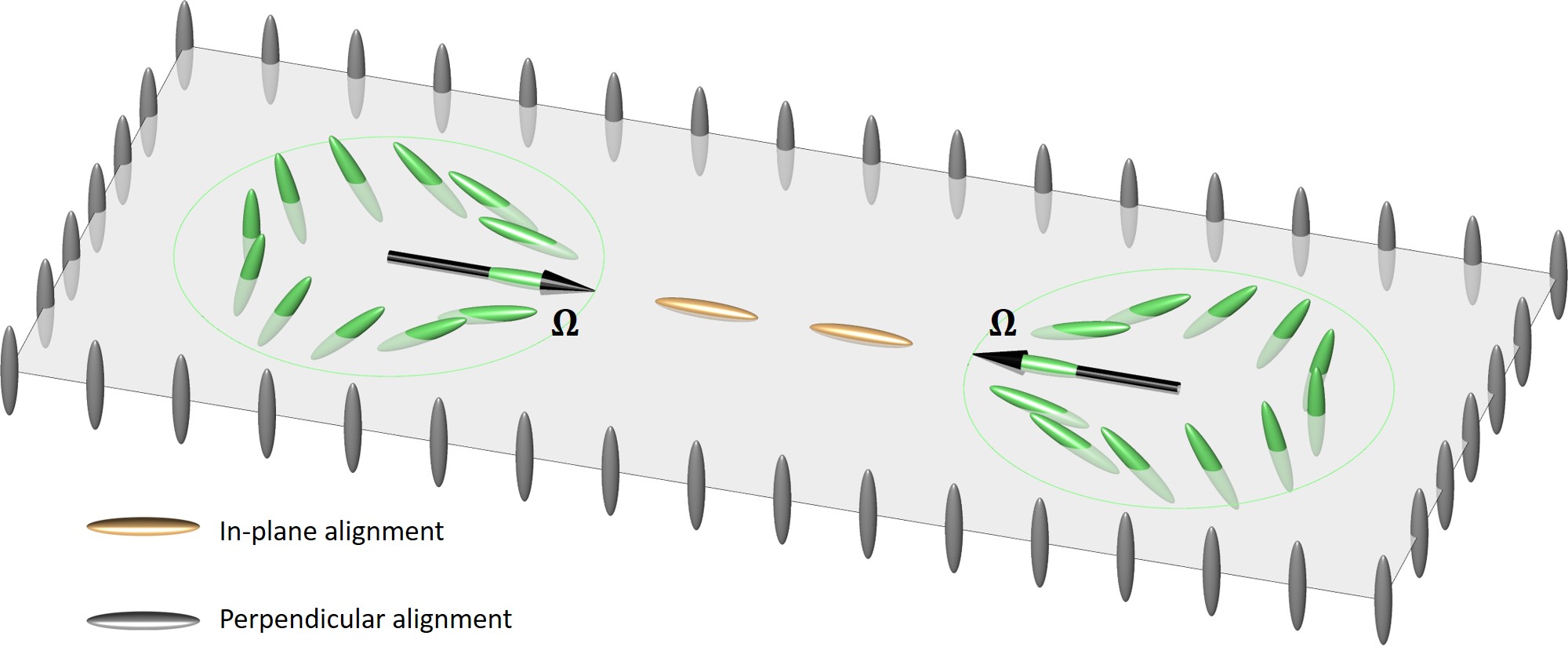}
	\caption{Adjacent twist-type surface defects (green) tend to align anti-parallel to each other in contractile systems to minimise the region of in-plane alignment. This results in lines of planar surface anchoring between the defects (orange directors).}
	\label{fig_twist_line}
\end{figure}

By contrast, contractile activity leads to a remarkable stripe pattern in the director alignment on the surface. While most parts of the surface show strong perpendicular (homeotropic) active alignment, this is interspersed with thin stripes of clear in-plane ordering (see Fig.~\ref{fig_act_anch}(b), Movie~S1 \citep{si}). The stripes start and terminate at surface defects which are mostly twist-type with $\beta \approx \pi/2$  (Fig.~\ref{fig_betasurf}). Wedge-type surface disclinations with $\beta = 0,\pi$ are suppressed as they would create a region of in-plane alignment in the vicinity of the defect. Twist-type disclinations on the other hand only introduce a small region of in-plane alignment along one specific direction (Fig.~\ref{fig_beta_sketch}) and are therefore favoured. Adjacent twist-type disclinations tend to align with their rotation vectors antiparallel, in a way that minimises the area of unfavourable in-plane alignment, thereby creating the observed stripe pattern (Fig.~\ref{fig_twist_line}).

Perfect perpendicular alignment at the surface would prevent disclination lines in the bulk terminating there as all disclination lines introduce some degree of in-plane alignment at the surface. Indeed, because topological charge is conserved, a sphere with perfect homeotropic alignment at the interface would force the system to form a $+1$ defect loop in the bulk \citep{binysh2020three, vcopar2019topology}. However, $+1$ defect loops in the bulk are associated with large elastic energy of the liquid crystal which usually cannot be overcome by active anchoring.

\subsection{\label{sec:defsurf}Disclination line dynamics in spherical droplets}

\begin{figure}
	\centering
	\includegraphics[width=8.6cm]{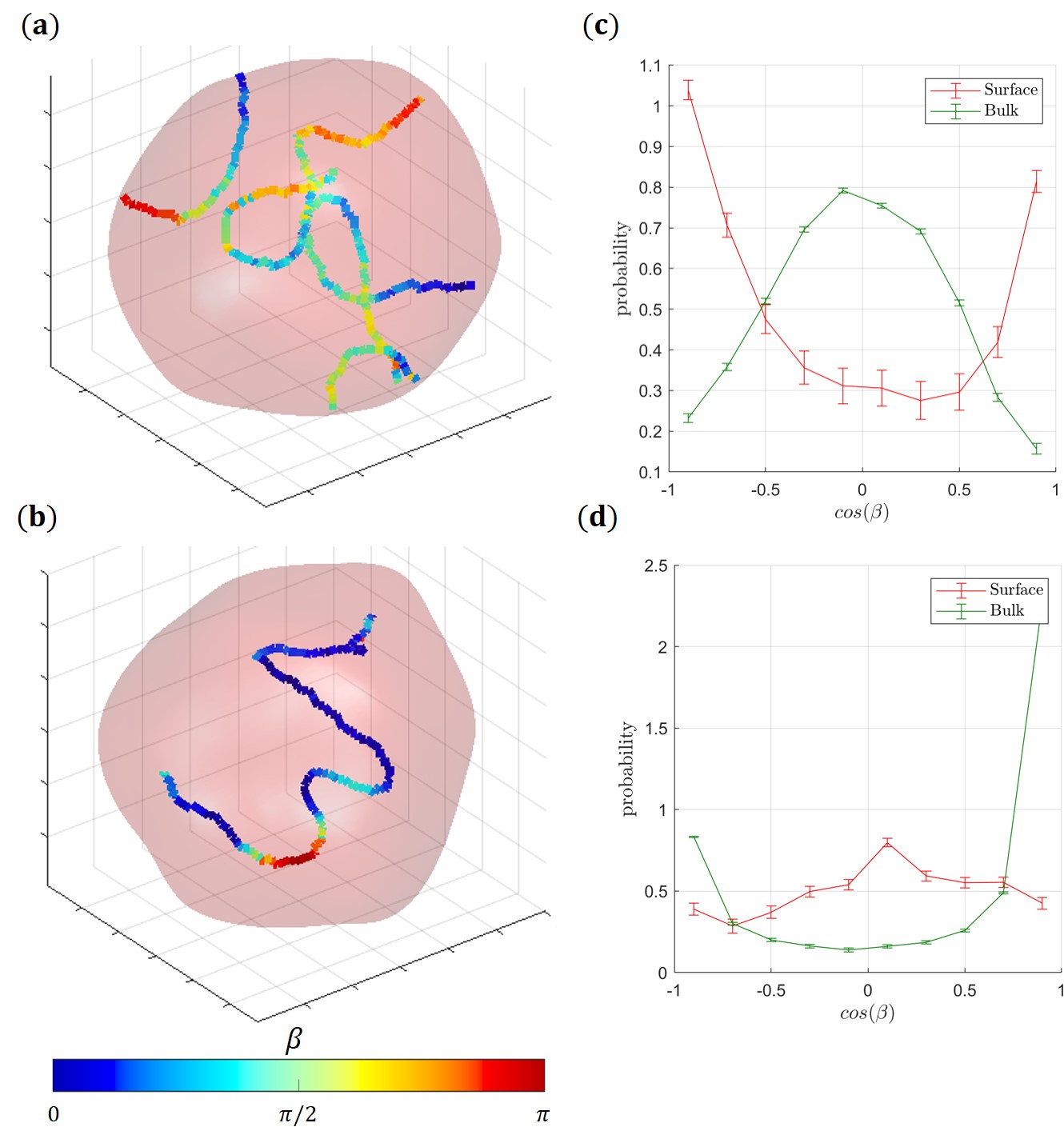}
	\caption{Snapshots of disclination lines in an (a) extensile and (b) contractile droplet. The colour bar shows the twist angle $\beta$ which characterises the director configuration which varies continuously along the line. (c) As a consequence of active anchoring, for extensile systems disclinations close to the surface tend to be wedge-type whereas those in the bulk tend to be twist disclinations as is evident from the distribution of $\beta$. (d) For contractile systems disclinations close to the surface slightly tend to be twist-type whereas those in the bulk tend to be wedge disclinations.}
	\label{fig_beta_drop}
\end{figure}

We now focus on disclination line dynamics inside spherical droplets which are essentially undeformed by active forces. In the regime of active turbulence, disclination lines constantly form and annihilate. Previous investigations of active turbulence in bulk systems found that the dominant excitations of three-dimensional active nematics are charge-neutral disclination loops which undergo complex dynamics and recombination events \citep{duclos2020topological}. In droplets however disclination lines do not need to form loops as they can also exist as growing or shrinking half circles or lines which terminate at the surface at positions $[\mathbf{x}_1$, $\mathbf{x}_2]$. The distance between endpoints (surface defects) of disclination lines $D=|\mathbf{x}_1 - \mathbf{x}_2|$ first scales linearly with the total length $L$ of lines before plateauing at a finite value as the lines increase in length. For disclination lines which are short compared to the droplet radius ($L<R$), the mean endpoint separation scales as $D\approx (2/\pi) L$, showing that the lines mainly nucleate or annihilate at the droplet's surface as half-circles. The endpoints of very long disclination lines ($L \gg R$) are randomly distributed over the droplet's surface so the mean separation of endpoints converges towards $D/R \approx 4/3$ (Fig.~S2 \citep{si}). 
The local properties of disclination lines can be classified by the twist angle $\beta$ which varies continuously along the line. We find that the director profiles of disclination lines close to the surface are heavily influenced by active anchoring which favours wedge- (twist)-type disclination lines for extensile (contractile) activity (Fig.~\ref{fig_beta_drop}, Movie~S1 \citep{si}). This is in contrast to the bulk of the drop where extensile (contractile) activity favours twist- (wedge)-type disclinations. The distributions of $\beta$ at the surface and in the bulk are compared for extensile and contractile droplets in Fig.~\ref{fig_beta_drop}(c) and Fig.~\ref{fig_beta_drop}(d) respectively.

\begin{figure}
	\centering
	\includegraphics[width=8.6cm]{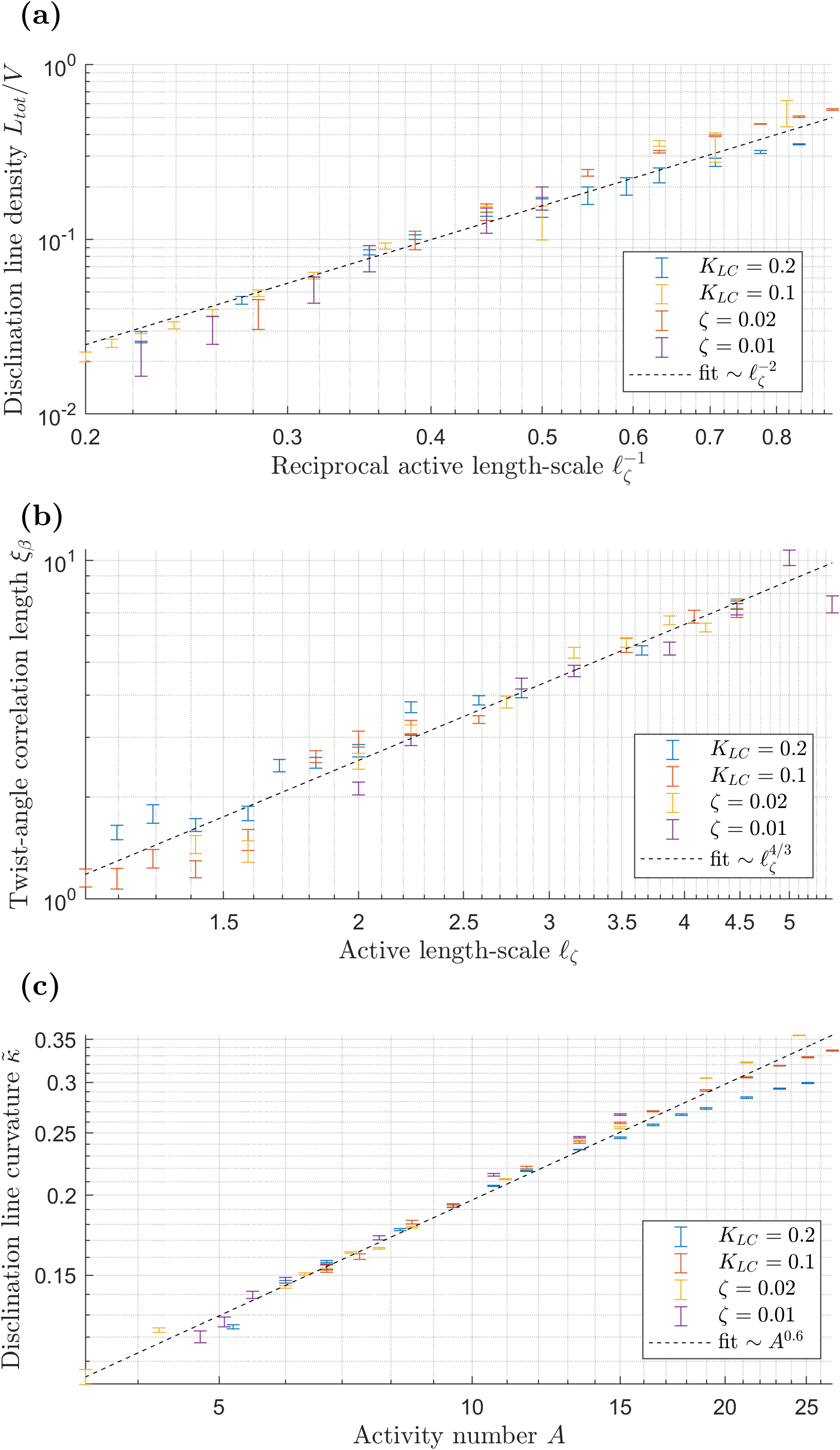}
	\caption{(a) Mean area density of disclination lines, defined as the total length $L_{tot}$ of all disclination lines divided by the volume $V$, as a function of the inverse active length scale $1/\ell_\zeta = \sqrt{\zeta/K_{LC}}$. (b) Persistence length  $\xi_\beta$ of the twist angle $\beta$ as a function of the active length-scale. (c) Mean curvature $\tilde{\kappa}$ of disclination lines as a function of activity number $A=R/\ell_\zeta$.}
	\label{fig_beta_scaling}
\end{figure}

Behind the apparent disorder of three-dimensional active turbulence we identify an {\it active length-scale} $\ell_\zeta \propto \sqrt{K_{LC}/\zeta}$ which controls the average disclination line density and coincides with the active length scale governing the decay of the vorticity correlations in two-dimensional active turbulence \citep{doostmohammadi2018active}. Fig.~\ref{fig_beta_scaling} shows the area density of disclination lines, 
defined as the total length $L_{\mbox{tot}}$ of all disclination lines divided by the
droplet volume V,  as a function of $K_{LC}$ and $\zeta$, confirming that it scales with the active length-scale as $ \ell_\zeta^{-2}$. Moreover, measuring the mean variation of twist angle $\beta$ along disclination lines shows an exponential convergence towards a plateau with a characteristic length scale $\xi_\beta$ (Fig.~S3 \citep{si}). We find that in the turbulent regime, the correlation length $\xi_\beta$ of twist angle $\beta$ scales with the active length-scale as $\xi_\beta  \propto \ell_\zeta^{4/3}$. The exponent $4/3>1$ reflects that disclination lines are curved structures in three-dimensional space, hence correlations do not decay linearly along lines. Based on these observations, we define the dimensionless {\it activity number} $A=R\sqrt{\zeta/K_{LC}}$ as the ratio of $droplet$ size $R$ to active length-scale $\ell_\zeta$. The mean curvature $\tilde{\kappa}$ of disclination lines increases with activity number roughly as $\tilde{\kappa} \sim A^{0.6}$. 

\begin{figure}
	\centering
	\includegraphics[width=8.6cm]{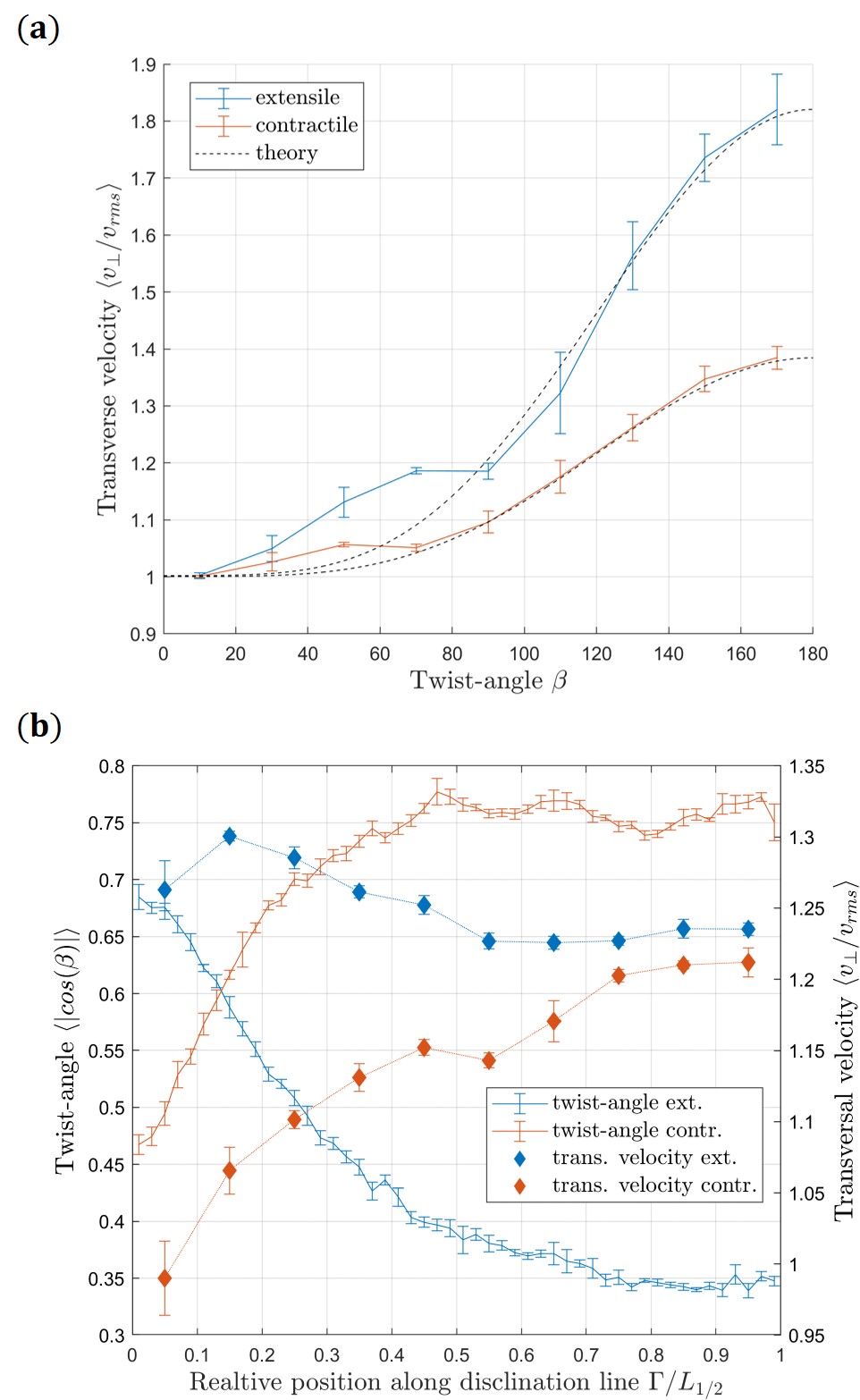}
	\caption{(a) Mean transverse velocity $\langle v_\perp \rangle$ of disclination line segments as a function of local twist angle $\beta$.  The transverse velocity is normalised by the average root mean square velocity $v_{rms}$ of the fluid. The fit is to the theoretical prediction \citep{binysh2020three}, eqn.~(\ref{eqn_spv}), with the constant of proportionality obtained from a least-square fit.
    (b) Mean twist angle $\langle |\cos(\beta)| \rangle$ and mean transverse velocity $\langle v_\perp \rangle$ as a function of the relative position along disclination lines, where values $0$ and $1$ denote positions close to the surface and in the bulk, respectively. The relative position is measured as the ratio of $\Gamma$, which is the path length between each point and the nearest end-point of the disclination line, and $L_{1/2}$, which is half the total line length.}
	\label{fig_beta_boltzmann}
\end{figure}

Due to activity disclination lines move through the fluid with an associated transverse self-propulsion velocity $v^{SP}_{\perp}$ (eqn.~\ref{eqn_spv}). In addition, they are passively advected by the surrounding flow and thus  the mean transverse velocity of disclination line segments is expected to follow $\langle v_{\perp} \rangle \approx v_{rms}+v^{SP}_{\perp}$, where $v_{rms}$ denotes the root-mean-square velocity of the fluid inside the droplet. 
Surprisingly this simplified model of uncoupled, self-propelled line segments closely matches the velocity profile we observe in turbulent droplets for large $\beta$ (Fig.~\ref{fig_beta_boltzmann}(a)). As disclination lines consist of line segments with varying self-propulsion velocities, more active line segments are elastically coupled with more passive segments. Passive line segments are thus pulled around by elastic interactions in addition to passive advection by the flow, which is the reason for the deviation of $\langle v_{\perp} \rangle$ from the theoretical prediction for small $\beta$.

We have argued that, as a consequence of active anchoring, the distribution of $\beta$ varies between disclination lines near the surface and those deep into the bulk (Fig.~\ref{fig_beta_boltzmann}(b), solid lines). This is mirrored in the average self-propulsion velocity (Fig.~\ref{fig_beta_boltzmann}(b), diamonds). In extensile systems the wedge-type defects close to the surface on average move faster than the twist-type line segments in the bulk. Similarly, contractile activity causes line segments in the bulk, which are mostly wedge-type, to move faster than surface defects.

\subsection{Extensile activity triggers the formation of finger-like protrusions}

\begin{figure}
	\centering
	\includegraphics[width=8.6cm]{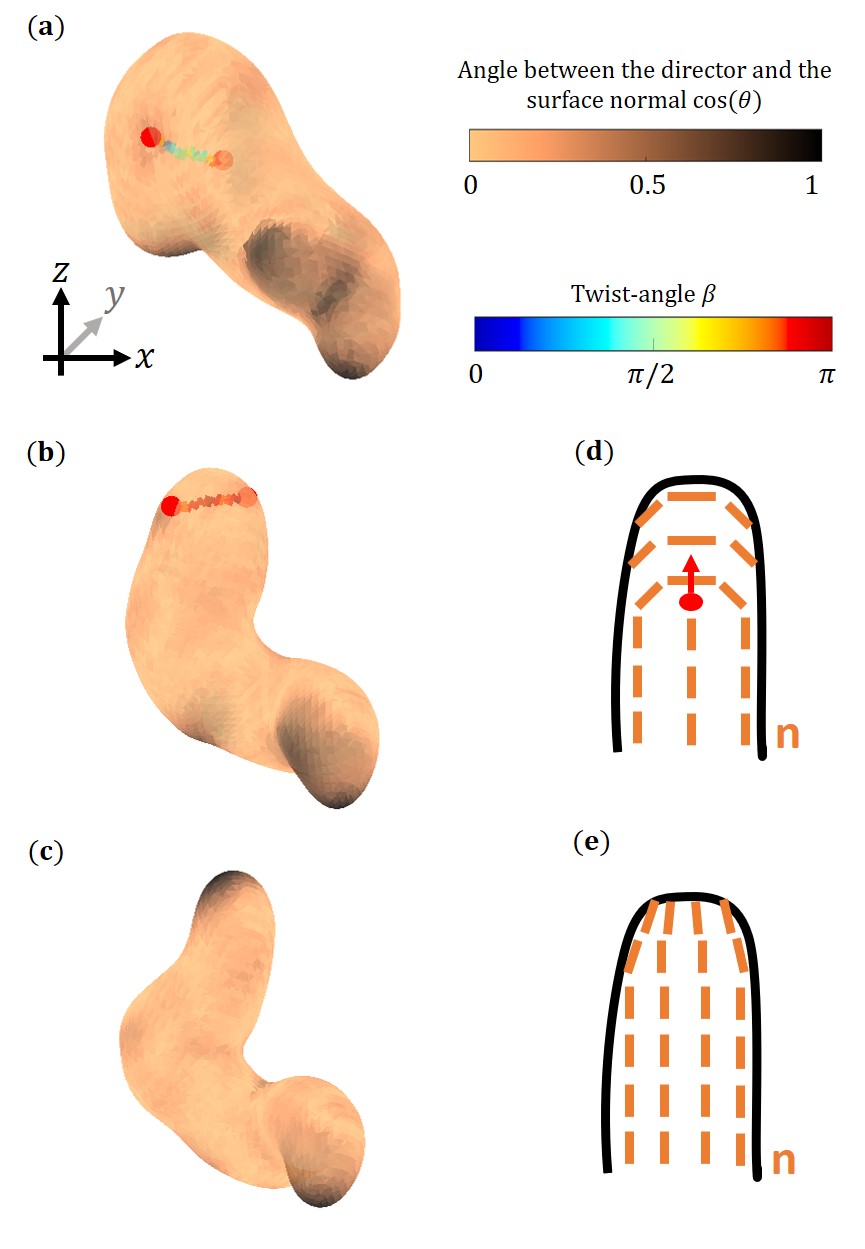}
	\caption{Formation of finger-like protrusions by motile disclination lines shown by three snapshots at times $t_1 < t_2 < t_3$. Panels (a)-(c) show the three-dimensional droplet shape with colour coding showing the local surface alignment of the director field. Disinclination line segments are coloured according to the twist angle $\beta$ with the same colour coding as in Fig.~\ref{fig_beta_drop}. (a) Disclination line with varying twist angle $\beta$ moves towards the interface. Active flow pushes the interface outwards creating a bulge along the self-propulsion direction of the disclination line. (b) The small protrusion width combined with in-plane alignment at the surface stabilises the disclination line into an almost straight configuration with $\beta \approx \pi$. (c) Disclination line moves out of the droplet leaving behind a defect-free protrusion with an aligned director field.	Panels (d),(e) show a schematic diagram of the director field and active forces. The red arrow in panel (d) denotes the self-propulsion direction of the $\beta \approx \pi$ disclination lines. As shown in panel (c), disclinations leave behind an area of perpendicular surface alignment (dark regions) at the end of protrusions which slowly retract due to surface tension and bending rigidity. Snapshots created from simulation with droplet size $R=15$, see also Movie~S2 \citep{si}.}
	\label{fig_protrusion}
\end{figure}

We now consider lower values of the surfaces tension such that the active flows are strong enough to deform the droplet, which leads to the formation of defect-mediated, finger-like protrusions for extensile activity. Unlike in stiff droplets, the active flow field around disclination lines close to the surface can push the interface outwards creating a bulge along the self-propulsion direction of a disclination line (Fig.~\ref{fig_protrusion}(a)). 
As the disclination line continues to move outwards towards the interface, the local bulge extends and forms a thin protrusion. 
Due to the reduced separation of interfaces compared to the spherical droplet, the properties of disclination lines inside protrusions are dominated by active anchoring which favours in-plane surface alignment. This causes disclination lines to span the width of the protrusion as a straight line with most line segments resembling wedge-type $+1/2$ defect profiles ($\beta \approx \pi$), thereby enhancing the directed self-propulsion of the disclination line even further. The $+1/2$ defect profile aligns along the protrusion axis as this configuration satisfies the in-plane anchoring condition everywhere at the surface (Fig.~\ref{fig_protrusion}(b,d)). This quasi-two dimensional movement of $+1/2$ defect lines has also been observed in active nematic films confined to a channel below a critical wall separation \citep{shendruk2018twist}. Wedge-type $-1/2$ defect profiles ($\beta \approx 0$) are not observed in protrusions as they are passive and lack the self-propulsion necessary to form a bulge in the first place.

The straight $+1/2$ defect line eventually reaches the tip of the growing protrusion and moves out of the nematic droplet, leaving behind a homogeneous director field which is aligned along the protrusion axis (Fig.~\ref{fig_protrusion}(c,e)). Thereby an area of perpendicular surface alignment is introduced at the end of protrusions, which are points of large negative mean curvature. The surface alignment $\cos(\theta)$ is therefore correlated to the local mean curvature of the interface (Fig.~S4 \citep{si}). In the absence of disclination lines mediating active forces, the aligned protrusions slowly retract due to surface tension and bending energy. The constant formation and retraction of droplet protrusions is also shown in Movie~S2 \citep{si}.

\begin{figure}
	\centering
	\includegraphics[width=8.6cm]{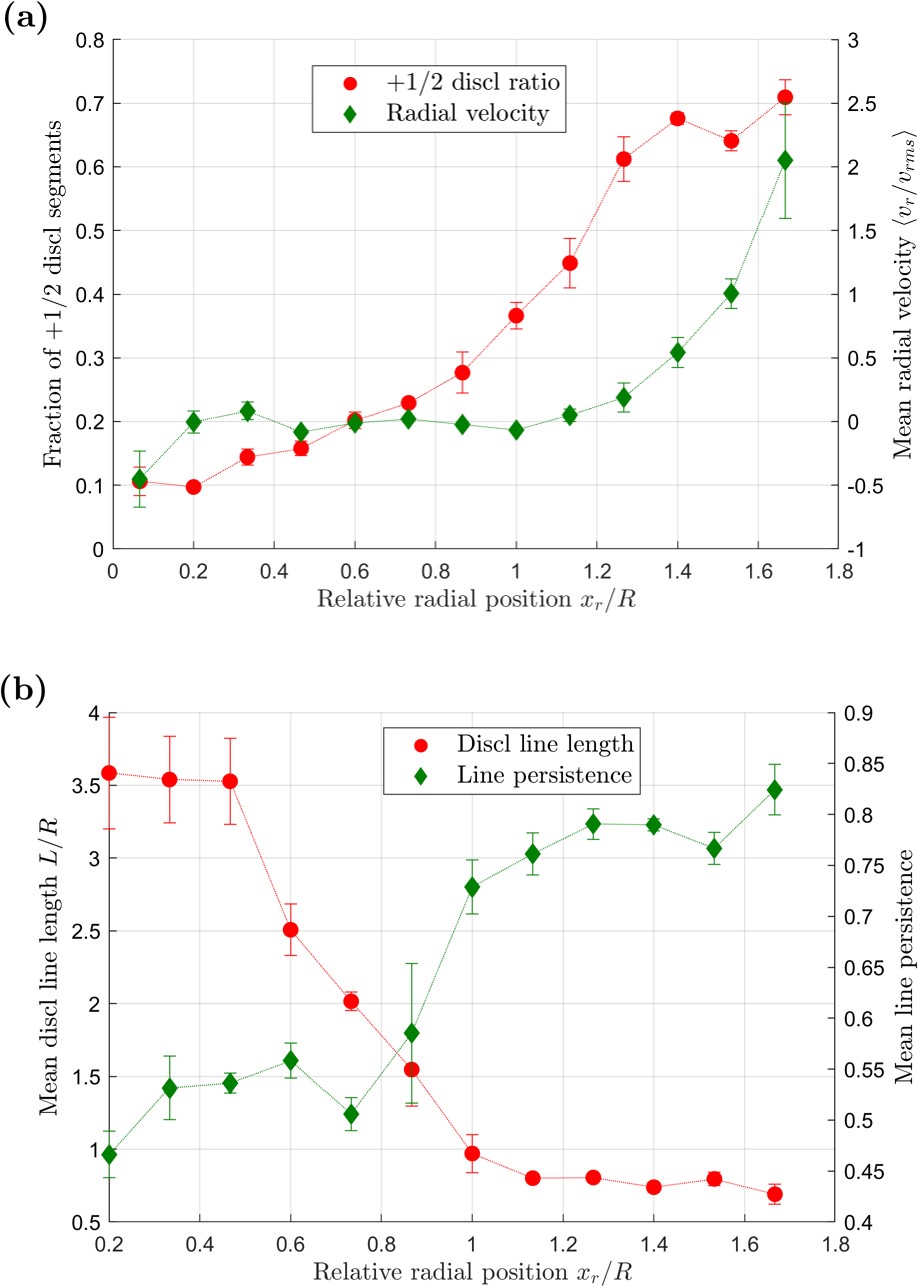}
	\caption{Disclination line properties for extensile activity as a function of radial position $x_r$ in an initially spherical droplet of radius $R$. (a) Most of the disclination lines in protrusions ($x_r > R$) show a $+1/2$ director configuration for which the self propulsion velocity $v_r$ points radially  outwards. (b) Disclination lines in protrusions ($x_r > R$) are nearly straight lines whose length is limited by the diameter of the protrusion. The radial component of the self propulsion velocity $v_r$ was obtained by taking the droplet's centre of mass as the reference point of a spherical coordinate system. The fraction of $+1/2$ disclination lines was defined as the fraction of disclination line segments with twist angle $3/4 \pi\leq\beta\leq\pi$ and the persistence of disclination lines as the ratio of endpoint distance over total line length.} 
	\label{fig_protrusion_properties}
\end{figure}

To further quantify the mechanism of protrusion formation we measured several properties of disclination lines as a function of radial position $x_r$ ,with the droplet's centre of mass being the reference point of a spherical coordinate system. For approximately ellipsoidal droplet shapes,  $x_r$ can be used as a proxy to divide the initially spherical droplet of radius $R$ into a bulk domain ($x_r<R$) and a protrusion domain ($x_r>R$).  Disclination lines in the bulk ($x_r<R$) of soft droplets undergo chaotic movement ($\langle \mathbf{v_r} \rangle \approx 0$) and are mostly twist type. By contrast, in protrusions ($x_r>R$) disclination lines mostly consist of $+1/2$ line segments ($\approx 70 \%$) and show persistent self-propulsion along the radial protrusion axis ($\langle \mathbf{v_r} \rangle \gg 0$, see Fig.~\ref{fig_protrusion_properties}(a)). Disclination lines in protrusions are nearly straight and their total length $L$ is limited by the width of the protrusion ($L(x_r)\approx \mbox{const}$ for $x_r>R$, see Fig.~\ref{fig_protrusion_properties}(b)).

Droplets of sizes much larger than the active length-scale $R \gg \ell_\zeta$ are strongly deformed and protrusions do not always grow along the radial axis (Fig.~S5(a) \citep{si}), thereby rendering the spherical approximation unsuitable. Still it can be observed that $+1/2$ defect lines are much more frequent in soft droplets with protrusions than in spherical droplets without protrusions (Fig.~S5(b) \citep{si}).

\begin{figure}
	\centering
	\includegraphics[width=8.6cm]{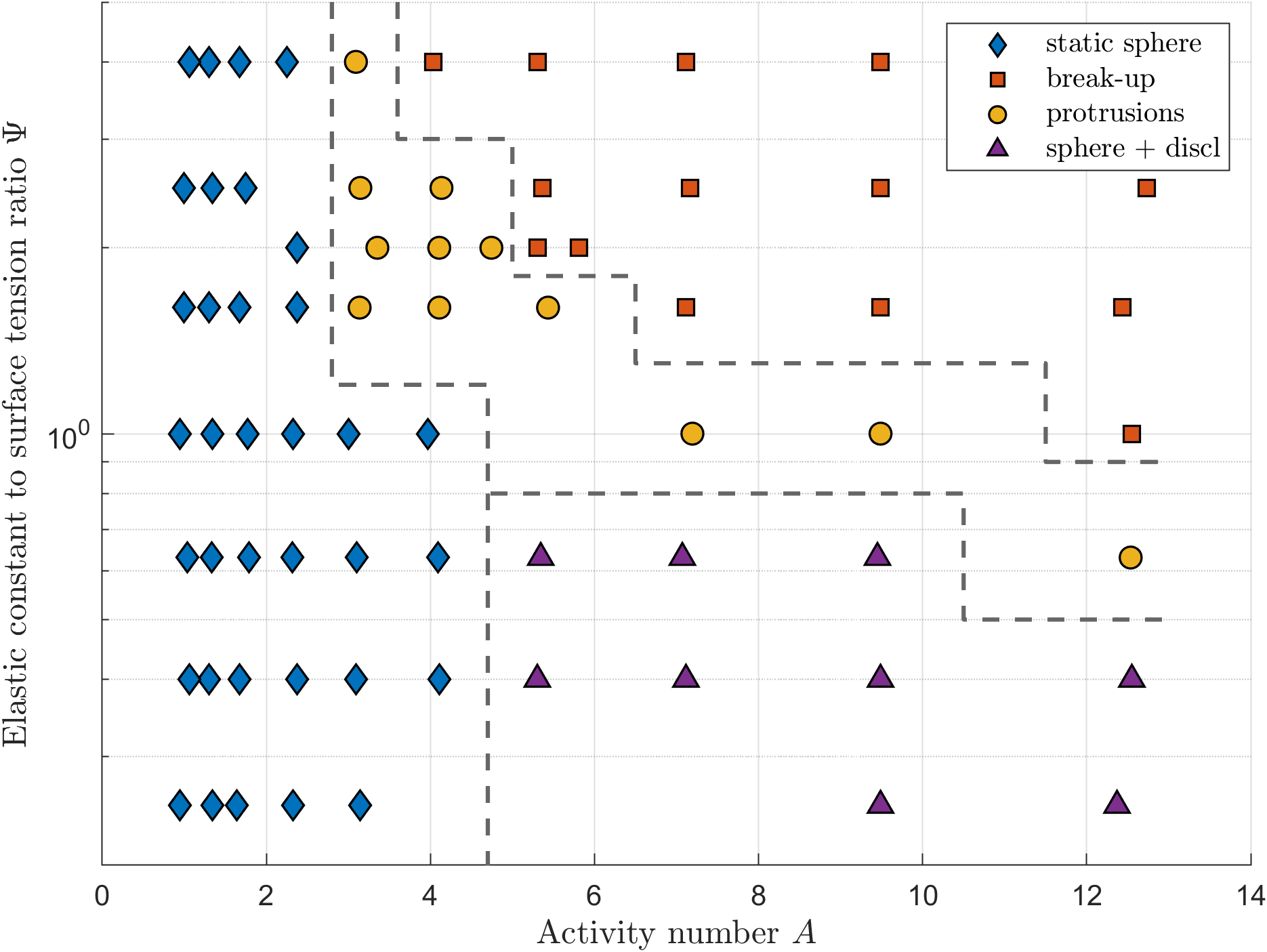}
	\caption{Morphology diagram of extensile droplets ($\zeta > 0$) as a function of the activity number $A=R \sqrt{\zeta /K_{LC}}$ and the elastic constant to surface tension ratio $K_{LC}/K_\varphi$. Parameters $\zeta$ and $K_{LC}$ were varied while the surface tension and droplet size were fixed to one of the following values: $K_\varphi \in \{0.1, 0.2\}$ $R \in \{15,30\}$.}
	\label{fig_ext_morph}
\end{figure}

The morphology of extensile droplets is determined by two control parameters: The activity number $A=R\sqrt{\zeta/K_{LC}}$ controls the density of disclination lines inside the bulk of droplets while the ratio of elastic constant to surface tension  $\Psi = K_{LC}/K_\varphi$ quantifies the energetic cost associated with nematic deformations in the bulk compared to deformations of the interface (assuming surface tension dominates the bending stiffness). The morphology of droplets as a function of $A$ and $\Psi$ is shown in Fig.~\ref{fig_ext_morph}. For very stiff interfaces ($\Psi \ll 1$), droplets are nearly spherical and host disclination lines if the activity $A$ is sufficiently large (blue diamonds, purple triangles). Soft interfaces ($\Psi \geq 1$) however allow the formation of finger-like protrusions as soon as disclination lines are formed inside droplets  (yellow circles). If active forces become much larger than the passive restoring force of the interface, active flows tear apart the droplet which breaks up into smaller parts (orange squares). The strength of protrusion formation can be quantified by the gyrification index, which is a function of $A$ and $\Psi$ (Fig.~S8(a) \citep{si}).

\subsection{Contractile activity triggers droplet invagination and surface wrinkles}

We now consider contractile activity which causes droplet invagination or the formation of comb-shaped droplet deformations creating wrinkle patterns on the surface. Unlike protrusion formation in extensile systems, droplet deformations in contractile systems originate from smooth director field deformations and are not mediated via disclination lines. 

\begin{figure}
	\centering
	\includegraphics[width=8.6cm]{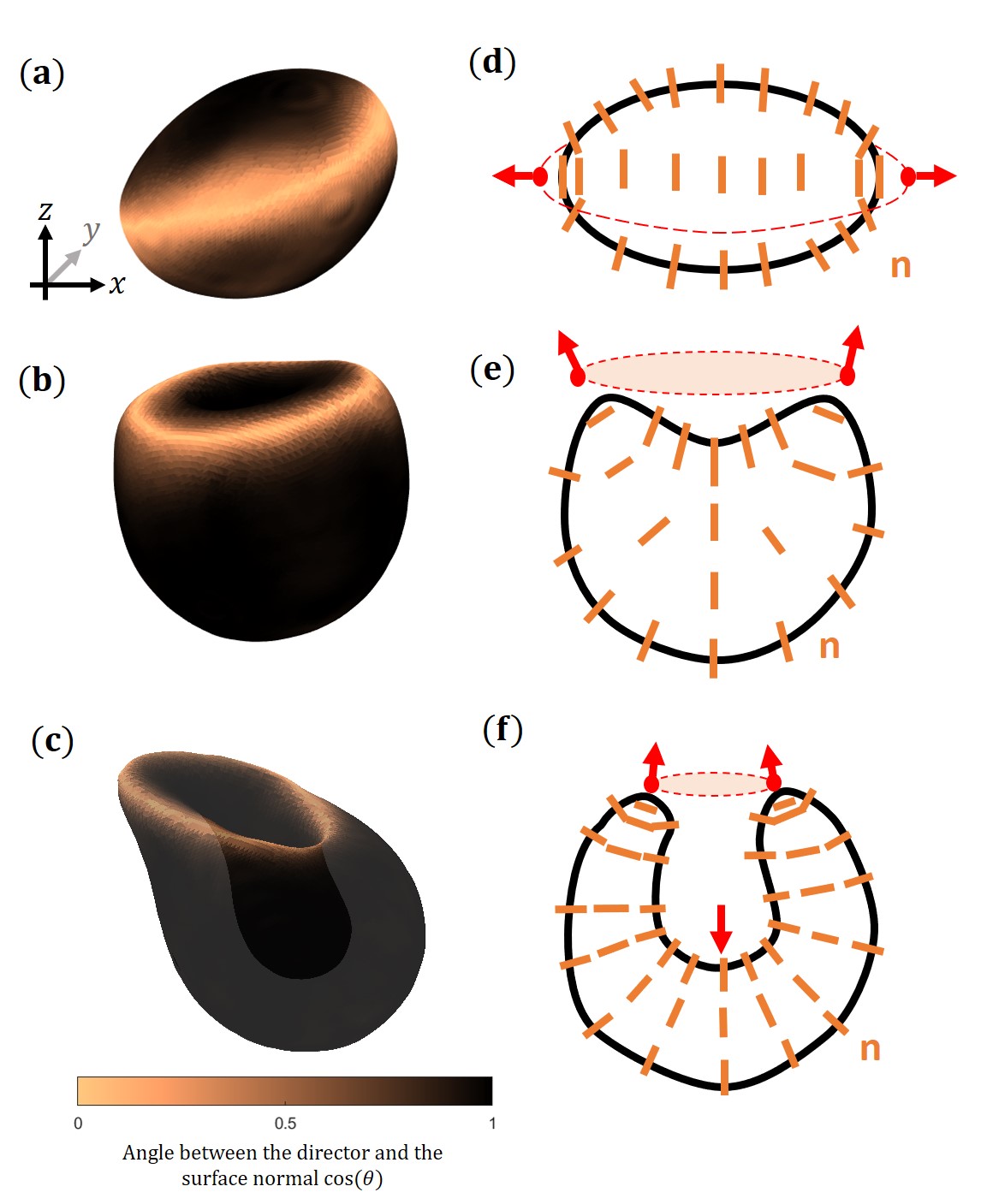}
	\caption{Initiation of droplet invagination by contractile activity shown by three snapshots at times $t_1 < t_2 < t_3$. Panels (a)-(c) show the three-dimensional droplet shape with colour coding indicating the local surface alignment of the director field. Panels (d)-(f) are schematic diagrams of the director field $\mathbf{n}$, showing the active forces (red arrows) and the position of the bend-ring (red, dotted circle). (d) The ring of in-plane alignment at the equator causes nematic bend deformations in the bulk. Contractile activity causes these bend deformations to push outwards (red arrows), thus deforming the droplet to an oblate shape. (e) While the ring of in-plane alignment moves towards one of the poles, the bend deformations in the bulk point perpendicular to the surface which causes the active forces to develop a component pointing towards the pole. This creates a dent in the centre of the ring. (f) Due to the perpendicular surface alignment caused by active anchoring, the dent in the centre of the ring leads to a director splay deformation in the bulk. Contractile activity causes this splay deformation to push inwards, thus increasing the depth of the dent. Snapshots created from simulation with elastic constant $K_{LC}=0.4$. 
	}
	\label{fig_v1_loop}
\end{figure}

In contractile systems active anchoring favours normal surface alignment. Every spherical or topologically equivalent surface with normal surface alignment everywhere enforces the formation of at least one $+1$ disclination-loop in the bulk due to topological constraints. These loops are usually associated with a large elastic energy cost due to strong deformations of the director field and thus are only observed in drops if strong normal anchoring is enforced by additional terms in the free energy. Otherwise, in the absence of disclination lines, the director field  forms a ring with in-plane surface alignment encircling the droplet (Fig.~\ref{fig_v1_loop}(a)) to maximize the area of perpendicular surface alignment favoured by active anchoring while avoiding the formation of a $+1$ defect-loop in the bulk. The ring of in-plane surface alignment is associated with nematic bend deformations in the bulk; thus we will refer to this structure as a {\it bend-ring} from now on (Fig.~\ref{fig_v1_loop}(d)). Contractile activity causes the bend-ring to push outwards, thereby deforming the droplet to an oblate ellipsoid until passive forces arising from surface tension and membrane rigidity counterbalance the active force.

If active forces are small compared to the passive restoring forces arising from membrane and director field deformations, the oblate bend-ring configuration is stable. Perturbations of ring shape and position are associated with increasing elastic deformations of the membrane and the director field, thereby creating an energy barrier which stabilises the configuration (Fig.~S6 \citep{si}). The exact free energy profile depends on model parameters such as the surface tension $K_\phi$, membrane rigidity $\kappa$, nematic elastic constant $K_{LC}$ and bulk properties. 

If active forces are sufficiently strong to overcome the passive restoring force, however, the bend-ring at the equator is unstable. It contracts while moving towards one of the poles and its motion results in invagination leading to a cup-shaped configuration of the droplet.

This instability arises from the active flow set up by the director configuration of the bend-ring which pushes outwards along the deformation axis (Fig.~\ref{fig_hydr_inst}(a)). Any small deviations from a perfect ellipsoidal droplet shape cause a component of the active flow to point towards one of the poles. The resulting droplet deformation rotates the axis of the bend-ring further towards the pole thereby in turn further increasing the active force component in the direction of the pole (Fig.~\ref{fig_v1_loop}(b,e)). 

As the contracting bend-ring approaches the pole, it creates an area of splay deformation in its centre due to active anchoring enforcing perpendicular surface alignment (Fig.~\ref{fig_v1_loop}(e)). This splay deformation sets up an active flow pointing inwards (Fig.~\ref{fig_hydr_inst}(b)). The outward pointing forces of the bend-ring together with the inward pointing forces in its centre combine to drive droplet invagination (Fig.~\ref{fig_v1_loop}(c,f)). If the magnitude of active stress $\zeta$ is not reduced after complete invagination, this process eventually leads to droplet break-up as the passive restoring forces resulting from membrane rigidity and surface tension are not sufficient to compensate the active forces. However, if the system is active only for a certain amount of time and $\zeta$ is reduced after the initial invagination, the cup-shaped drop can be stabilised, analogous to the natural course of morphogenetic events, which are controlled by biochemical signals and begin and end at a predefined time \citep{rozman2020collective}.

\begin{figure}
	\centering
	\includegraphics[width=8.6cm]{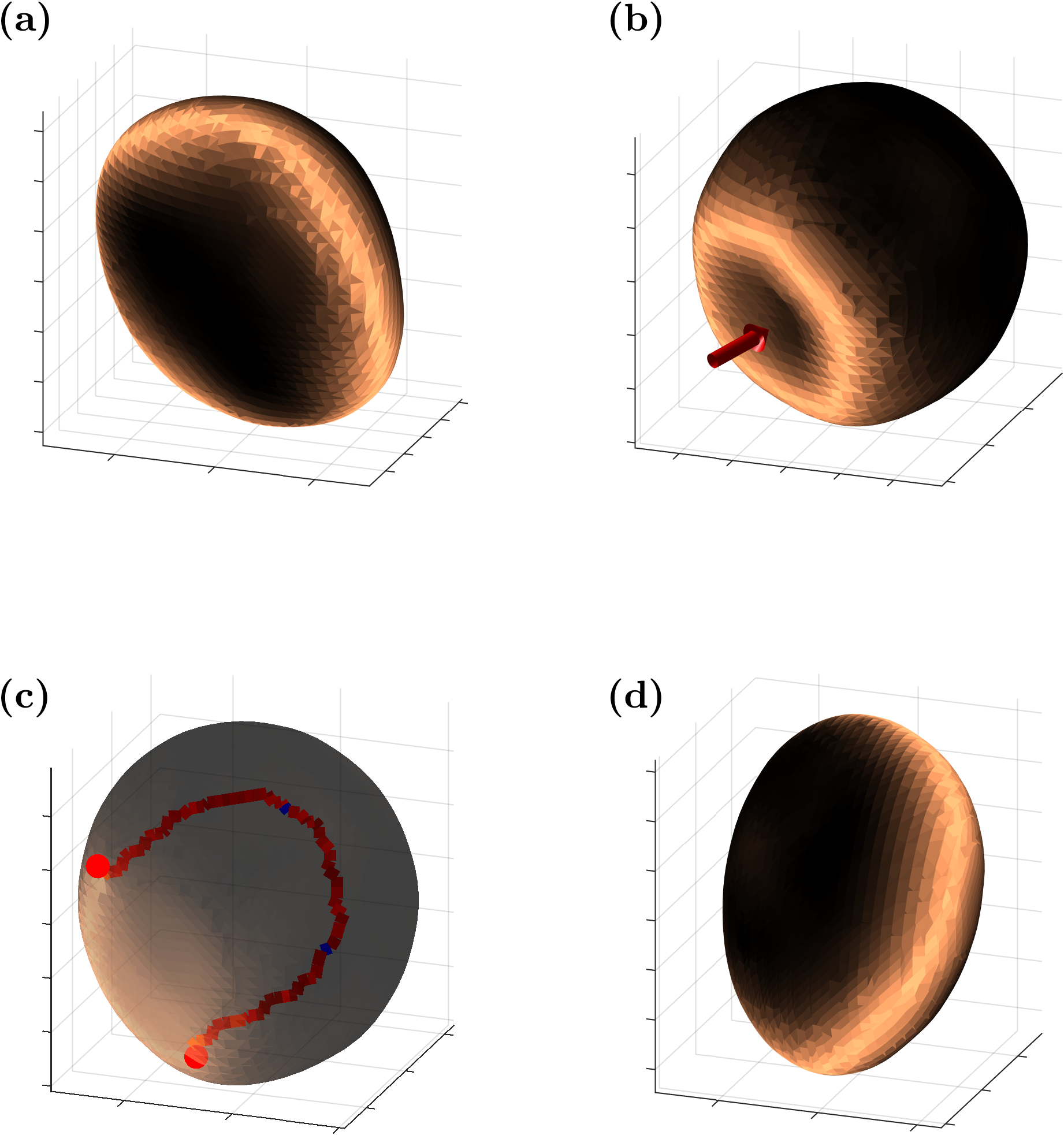}
	\caption{Droplets perform an {\it run-and-tumble} motion if the interface is too stiff for active forces to cause complete droplet invagination. Instead, a bend stripe at the equator deforms the droplet to an oblate shape (a), which then shrinks, forming a dip at the centre of the bend-ring where a splay deformation in the bulk pushes the droplet forward (b). Then the elastic deformations relax via the nucleation of a $\beta=\pi$ half-loop (c). The self-propulsive disclination line moves through the droplet and eventually reaches the interface. Thereby the bend-ring is re-oriented along a random direction and the process repeats (d). Snapshots created from simulation with droplet size $R=15$, see also Movie~S3 \citep{si}.}
	\label{fig_act_RW}
\end{figure}

Alternatively, if the interface is too stiff for active forces to initiate drop invagination, droplets perform an active {\it run-and-tumble} motion (Fig.~\ref{fig_act_RW}, Movie~S3 \citep{si}). Instead of a large cavity, only a small dip is formed at the centre of the bend-ring where the interface resists further deformations and the splay configuration of the director field leads to flows which push the droplet forwards. Eventually, the director deformations relax via the nucleation of a $\beta=\pi$ half-loop and the self-propulsive disclination line moves through the droplet until it reaches the opposite interface. This re-orientates the bend-ring  and the process repeats along a random direction.

\begin{figure}
	\centering
	\includegraphics[width=8.6cm]{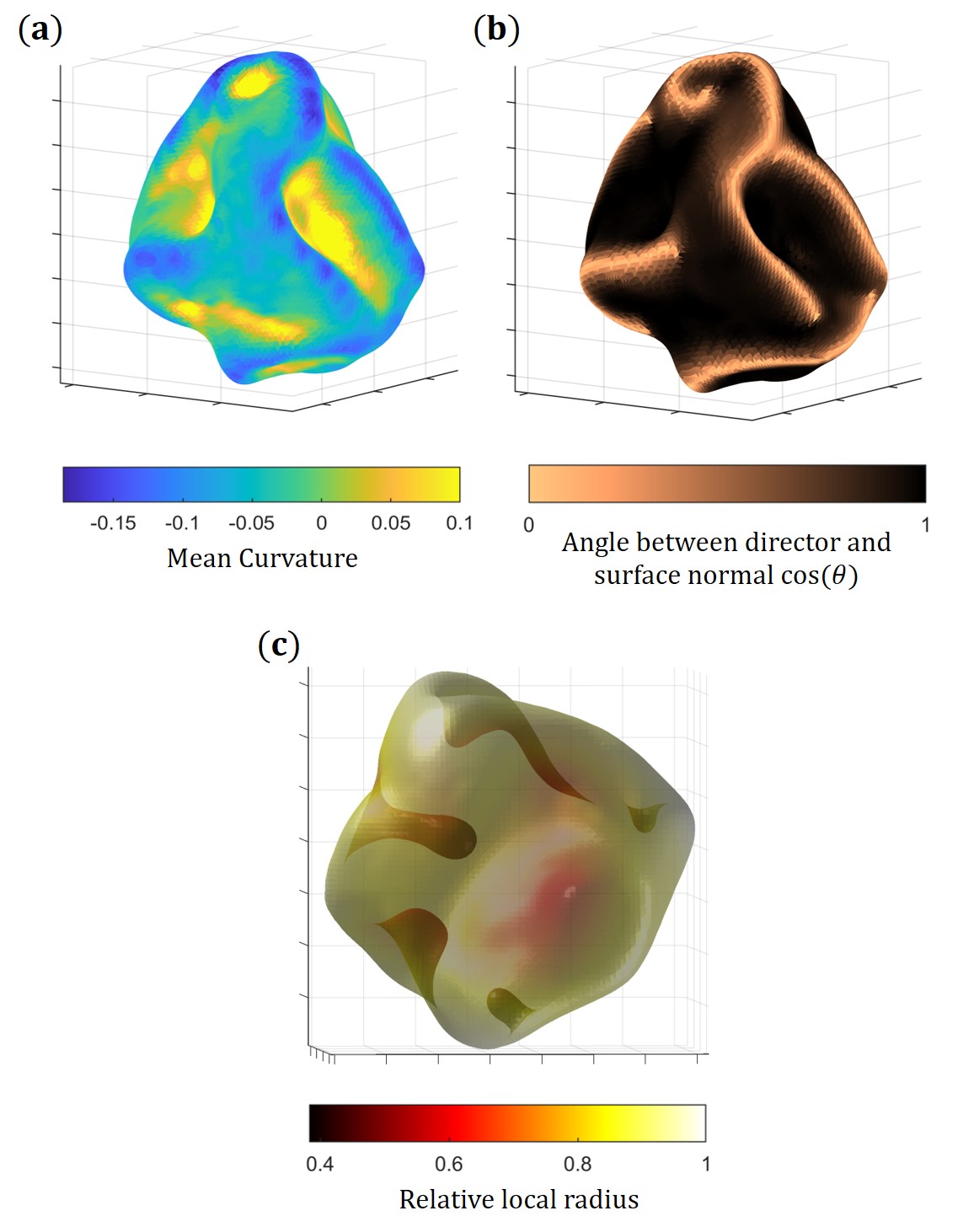}
	\caption{Snapshots of a contractile drop showing (a) the mean curvature and (b) the orientation of the director field on the surface. Contractile droplets which contain disclination lines show stripes of in-plane director alignment on the surface which connect endpoints of the disclination lines in the bulk (Fig.~\ref{fig_act_anch}). Bend and splay deformations of the director field near the surface create comb-shaped deformations along the stripes of in-plane alignment, resulting in a wrinkle pattern. In between stripes of in-plane alignment, dimples and valleys are driven into the drop as shown in snapshot (c) which depicts the local radius on a semi-transparent drop. The dynamical structure of active surface wrinkles is shown in Movie~S4 \citep{si}, together with a 3D view of the surface structure.}
	\label{fig_act_wr}
\end{figure}

In contractile droplets which are larger than the active length-scale $R \gg \ell_\zeta$, many disclination lines are present in the bulk and the surface shows characteristic stripes of in-plane director alignment connecting the endpoints of disclination lines (Fig \ref{fig_act_anch}). The stripes on the surface are associated with nematic bend deformations in the bulk which induce active forces pushing outwards. This creates comb-shaped deformations along the stripes of in-plane alignment, resulting in a surface wrinkle pattern which we will term {\it active wrinkling} (Fig \ref{fig_act_wr}, Movie~S4 \citep{si}). In addition the droplet deformation causes the perpendicularly aligned surface areas between in-plane stripes to form splay deformations in the bulk, which cause inward pushing, active forces that create dimples and valleys, further enhancing the wrinkle pattern. Along the in-plane ridges the mean curvature is negative while in the centre of holes it is positive. Surface alignment $\cos(\theta)$ is therefore correlated to the local mean curvature of the interface with in-plane surface alignment ($\cos(\theta)\approx 0$) associated with points of negative mean curvature and perpendicular surface alignment ($\cos(\theta)\approx 1$) predominating at points of positive mean curvature (Fig.~S7 \citep{si}).

\begin{figure}
	\centering
	\includegraphics[width=8.6cm]{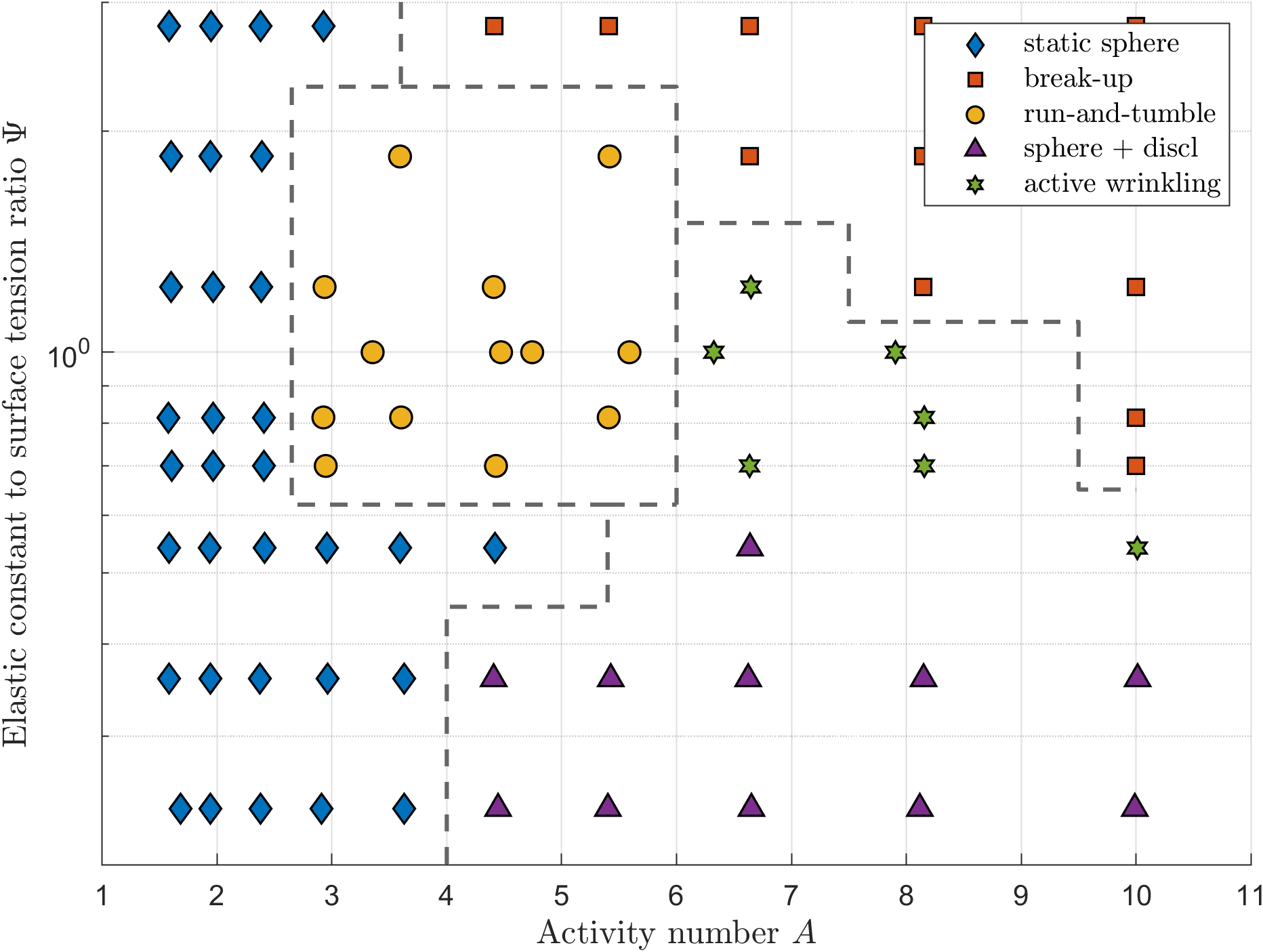}
	\caption{Morphology diagram of contractile droplets ($\zeta < 0$) as a function of the activity number $A=R \sqrt{\zeta /K_{LC}}$ and elastic constant to surface tension ratio $K_{LC}/K_\varphi$. There is a smooth transition between spherical droplets with disclination lines and active wrinkling. Parameters $\zeta$ and $K_{LC}$ were varied while the surface tension and droplet size were fixed to one of the following values: $K_\varphi \in \{0.1, 0.2\}$ $R \in \{15,30\}$.}
	\label{fig_contr_morph}
\end{figure}

As for the extensile case, the morphology of contractile droplets is determined by the activity number $A=R\sqrt{\zeta/K_{LC}}$ and ratio of elastic constant to surface tension $\Psi = K_{LC}/K_\varphi$ (Fig.~\ref{fig_contr_morph}). Very stiff interfaces ($\Psi \ll 1$) cause droplets to be nearly spherical and host disclination lines if activity $A$ is sufficiently large (blue diamonds, purple triangles). If interfaces are sufficiently soft ($\Psi \approx 1$), droplets can be either be static, oblate ellipsoids without disclination lines (blue diamonds), perform an run-and-tumble motion (yellow circles) or form active surface wrinkles (green stars). 
For very soft interfaces ($\Psi \gg 1$), droplets are static ellipsoids at low activity and full droplet invagination takes place at larger activity, leading the droplet to eventually break-up (orange squares). The degree of invagination and wrinkling can also be quantified by a gyrification index, which varies with $A$ and $\Psi$ (Fig.~S8(b) \citep{si}).

\section{Discussion}

In this paper we have investigated the morphology and defect dynamics in active deformable droplets in three dimensions, considering both extensile activity, as present in experimental systems such as MT–kinesin mixtures \citep{sanchez2012spontaneous}, human bronchial epithelial cells \citep{blanch2018turbulent} and Madine–Darby canine kidney (MDCK) cells \citep{saw2017topological}, as well as contractile activity, which is found in systems such as mouse fibroblast cells \citep{duclos2017topological} or actomyosin gels \citep{schuppler2016boundaries}. We have shown that active stresses cause in-plane (extensile) or perpendicular (contractile) surface-alignment of the director field at the interface of droplets. Unlike thermodynamic surface anchoring, which involves an anchoring free energy, {\it active anchoring} is a hydrodynamic effect due to the flows driven by gradients of the nematic ordering $\mathbf{Q}$ at the interface. Active anchoring has important consequences for the dynamics of disclination lines close to the interface, such as the preferential formation of wedge-like surface disclinations in extensile systems or the creation of in-plane aligned stripes connecting twist-like surface disclination in contractile systems. This in turn triggers the formation of finger-like protrusions in soft, extensile droplets or the creation of surface wrinkles and droplet invagination in contractile systems. 
Furthermore, we applied the ideas of 2D active turbulence, such as defect velocities and the existence of an active length-scale, to three dimensions. We identified an active length-scale $\ell_\zeta$ of three-dimensional active turbulence, which controls both the density of disclination lines and the correlation length of the twist angle $\beta$ along disclination lines in the bulk of nematic droplets.

Recent work has interpreted several 2D biological systems including bacteria biofilms \citep{yaman2019emergence,dell2018growing}, epithelial tissue \citep{saw2017topological}, spindle-shaped cell monolayers \citep{duclos2017topological} and actin fibers in regenerating Hydra \citep{maroudas2020topological} in terms of the theories of active nematics. It is interesting to ask whether similar ideas may prove useful in 3D to describe biological systems in which nematic constituents, such as protein filaments, eukaryotic cells or bacteria, are organised as solid, 3D structures with a confining interface. There are several examples where cells migrate collectively as a cohesive group, thereby maintaining supracellular properties such as collective force generation and tissue-scale hydrodynamic flow \citep{shellard2019supracellular, friedl2009collective, he2014apical}.
For example, in some modes of collective cell movement implicated in cancer invasion, a blunt bud-like tip consisting of multiple cells that variably change position and lack well-defined leader cells protrudes from the tumour \citep{friedl2011cancer, ewald2008collective}. Hence, the natural appearance of finger-like protrusions in active nematic droplets could serve as a new approach to explain collective cell invasion of tumors.

In eukaryotic cells, membrane shapes under mechanical stress are mostly controlled by the mechanics of the cortical actin cytoskeleton underlying the cell membrane which produces contractile stresses. It has been observed that membrane shapes mainly depend on the actin thickness, where thin shells show a cup-shaped deformation and thick shells produce membrane wrinkles \citep{kusters2019actin}. Our work provides a possible link between contractile activity and  the emergence of cell-surface ruffles, circular dorsal ruffles (CDRs) and caveolae \citep{buccione2004foot, hoon2012functions, echarri2015caveolae} at the surface of cells.
Surface invagination which plays a role in active transport through cell membranes might also be related to local contractile forces. During macropinocytosis (fluid endocytosis), extracellular fluid is brought into the cell through an invagination of the cell membrane forming a small vesicle inside the cell. Vesicles form from cell-surface ruffles that close first into open cups (ruffle closure) and then into intracellular vesicles (cup closure) \citep{swanson2008shaping, doherty2009mechanisms}, which is reminiscent of the dimples that are driven into contractile drops. Some morphologies observed in contractile drops, such as partial invagination and run-and-tumble motion are known to also occur in active polar droplets \citep{tjhung2017contractile, tjhung2012spontaneous}. In contrast to nematic systems, which are characterised by a headless director field $\mathbf{n}\leftrightarrow -\mathbf{n}$, polar systems are described by a polarisation vector $\textbf{p}$. As a result, polar models do not exhibit $\pm 1/2$ topological defects, which have been shown to contribute to the dynamics and flows in cell cultures and bacterial biofilms \citep{duclos2017topological, blanch2018turbulent, saw2017topological,dell2018growing, yaman2019emergence}. 

Up to now, most of the research on three-dimensional active nematics has been focused on the topology and dynamics of closed disclination loops in the bulk or disclination lines in fixed geometries \citep{binysh2020three, vcopar2019topology, duclos2020topological,carenza2019rotation,carenza2020chaotic}. Previous investigations of two-dimensional active nematics which are confined to the surface of a 3D shell explored the connection between 2D topological defects and shell morphology \citep{metselaar2019topology}. With this work we expanded the current understanding of three-dimensional active nematics, considering the interplay between motile disclination lines, deformations of the nematic director field and the droplet interface.

Future directions include the consideration of more complex geometries, in particular thick, nematic shells or the addition of an anchoring free energy enforcing a given surface alignment. The motivation behind this is that most biological systems during embryonic development form sheets or shells rather than solid structures. 
Another aspect when considering cell aggregates such as embryos or tumors is that they are often composed of different cell types with different mechanical properties and cellular responses \citep{marusyk2010tumor}. It would be interesting to extend our model to two or more distinct, nematic fluid phases with different mechanical properties and activity coefficients, thereby incorporating cellular heterogeneity within cell aggregates. Another important aspect is that cells are usually surrounded by a three-dimensional dense network of macromolecules, the extracellular matrix, which provides structural support and is known to strongly affect cell migration behavior \citep{ehrbar2011elucidating, pathak2012independent}. The investigation of how a viscoelastic medium or a dense polymer solution affects active droplets therefore appears to be a promising direction.

\begin{acknowledgments}
We would like to thank Kristian Thijssen for helpful discussions. This
project was funded by the European Commission’s Horizon 2020 research and innovation
programme under the Marie Sklodowska-Curie grant agreement No 812780.
\end{acknowledgments}

\end{document}